\documentclass[11pt]{article}
\usepackage{jmlr2e-mod}
\pdfoutput=1

\usepackage{color}

\usepackage{graphicx}
\graphicspath{{./images/}}
\usepackage{pslatex}
\usepackage{mathptmx}	
\usepackage{url}
\usepackage[table]{xcolor}
\usepackage{tabularx}
\usepackage{ltablex}
\usepackage{booktabs}
\usepackage{adjustbox}
\usepackage{pgffor}
\usepackage{float}
\usepackage{caption}
\usepackage{ifthen}
\usepackage{url}

\usepackage{algorithmic}
\usepackage{xcolor}
\usepackage{subcaption}
\usepackage{stmaryrd}
\usepackage[inline]{enumitem}
\usepackage{comment}

\usepackage{bm}

\usepackage{amssymb}
\usepackage{amsmath}
\usepackage[mathscr]{euscript}

\usepackage[linesnumbered,ruled]{algorithm2e} 
\usepackage{parskip}
\usepackage[absolute]{textpos}

\graphicspath{{./}}

\title{Analyzing the Performance Portability of Tensor Decomposition}

\author{\name S. Isaac Geronimo Anderson \email sgeroni@sandia.gov, igeroni3@uoregon.edu\\
\name Keita Teranishi \email knteran@sandia.gov \\
\name Daniel M. Dunlavy \email dmdunla@sandia.gov \\
\name Jee Choi \email jeec@uoregon.edu \\
\addr Sandia National Laboratories, Albuquerque, NM 87123, USA \\
\addr Sandia National Laboratories, Livermore, CA 94551, USA \\
\addr University of Oregon, Eugene, OR 97403, USA}

\begin{document}

\maketitle

\renewcommand*{\thefootnote}{(\fnsymbol{footnote})}
\renewcommand*{\thefootnote}{\arabic{footnote}.}
	
\begin{abstract}
We employ pressure point analysis and roofline modeling to identify performance bottlenecks and determine an upper bound on the performance of the Canonical Polyadic Alternating Poisson Regression Multiplicative Update (CP-APR MU) algorithm in the SparTen software library.
Our analyses reveal that a particular matrix computation, $\bm\Phi^{(n)}$, is the critical performance bottleneck in the SparTen CP-APR MU implementation. Moreover, we find that atomic operations are not a critical bottleneck while higher cache reuse can provide a non-trivial performance improvement.
We also utilize grid search on the Kokkos library parallel policy parameters to achieve 2.25x average speedup over the SparTen default for $\bm\Phi^{(n)}$ computation on CPU and 1.70x on GPU.
We conclude our investigations by comparing Kokkos implementations of the \emph{STREAM} benchmark and the matricized tensor times Khatri-Rao product (MTTKRP) benchmark from the Parallel Sparse Tensor Algorithm (PASTA) benchmark suite to implementations using vendor libraries. We show that \emph{with a single implementation} Kokkos achieves performance comparable to hand-tuned code for fundamental operations that make up tensor decomposition kernels on a wide range of CPU and GPU systems.
Overall, we conclude that Kokkos demonstrates good performance portability for simple data-intensive operations but requires tuning for algorithms with more complex dependencies and data access patterns.
\end{abstract}


\begin{figure}[b!]
\centering
\includegraphics[width=\textwidth]{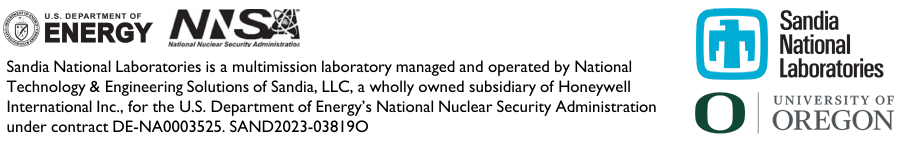}
\end{figure}

\clearpage
\section*{Acknowledgment}
We thank Richard Barrett, Rich Lehoucq, and D.S. Hollman of Sandia National Laboratories for providing useful suggestions associated with the work presented here. This work utilized the University of Oregon's high performance computer, Talapas.

\tableofcontents
\listoffigures
\listoftables
\clearpage

\section{Introduction}
\label{sec:intro}
\begin{figure}[!ht]
\centering
\includegraphics[width=.8\textwidth]{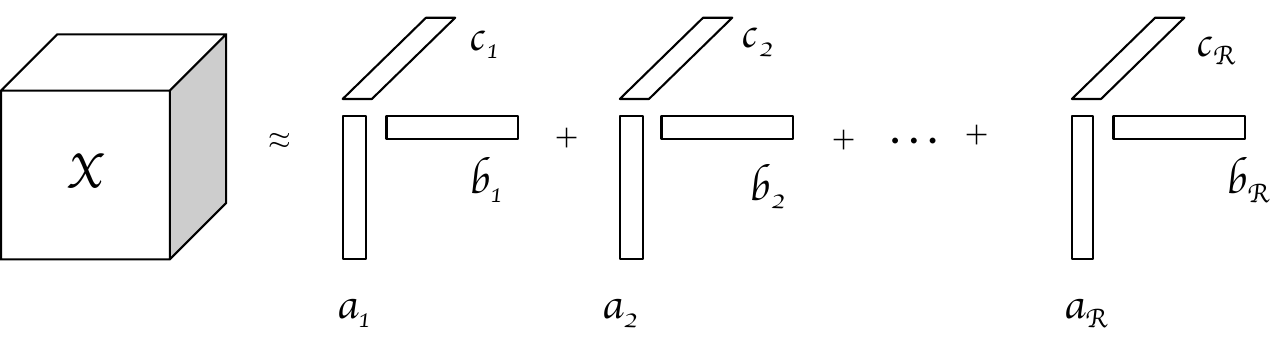}
\caption{Canonical polyadic decomposition of a 3-way tensor approximates a tensor by the sum of $R$ rank-one tensors, where each rank-one tensor is formed by the outer product of three vectors (e.g., $a_{1}$, $b_{1}$, and $c_{1}$), one for each mode (or dimension) of the tensor.}
\label{fig:cpd}
\end{figure}

Sparse tensor decomposition is a useful tool for extracting latent information from multiway data arising in many real-world applications, including healthcare~\cite{ho2014marble,He2019}, signal processing~\cite{sidiropoulos2017tensor}, cybersecurity~\cite{Fanaee-T2016,Bruns-Smith2016}, and more.
The Canonical Polyadic Alternating Poisson Regression (CP-APR) algorithm~\cite{doi:10.1137/110859063} using the multiplicative update (MU) method specifically targets \emph{sparse count data} and approximates the original tensor using the canonical polyadic decomposition (CPD) model, where a tensor is represented by a sum of rank-one tensors~\cite{KoBa09}.
The CPD model of a 3-way tensor is shown in Figure~\ref{fig:cpd}.
The vectors along each mode (e.g., $a_{1}$, $a_{2}$, ..., $a_{R}$ for mode-$1$) are combined to generate a factor matrix, where each vector is a column vector in the factor matrix.
For example, for a 3-way tensor of size $I_{1}\times I_{2}\times I_{3}$, CP-APR calculates three factor matrices $A\in\mathbb{R}^{I_{1}\times R}$, $B\in\mathbb{R}^{I_{2}\times R}$, and $C\in\mathbb{R}^{I_{3}\times R}$ for a rank-$R$ decomposition.

With the emergence of drastically different parallel architectures, {\it performance portability}---the ability to run the same program with little or no modification across different
architectures at an acceptable level of performance~\cite{exascale_sc, pp.org}---is critical in achieving optimal productivity on heterogeneous computing systems. 
It is impractical to write a high-performance software implementation for \emph{every} new architecture, and as such, performance-portable programming models will play an increasingly important role in both large-scale data analytics and high-performance computational science.

In this study, we employ the Pressure Point Analysis and Roofline Modeling techniques to identify key performance bottlenecks and determine an upper bound on the performance of the CP-APR MU algorithm implemented in the SparTen library~\cite{doi:sparten}.
SparTen leverages the Kokkos Core library ~\cite{6805038} to provide a performance-portable implementation of CP-APR that can be deployed in any Kokkos-supported hardware platform with a single implementation.
Our primary focus is on determining the viability of using Kokkos to provide efficient, performance-portable parallel computation support for the CP-APR MU algorithm.


The CP-APR MU algorithms is shown in Algorithm~\ref{alg:MU}. Note that this is an aggregation of the algorithm that was original presented in multiple parts when introduced by Chi and Kolda~\cite{doi:10.1137/110859063}. We note that the computation of the matrix $\bm\Phi^{(n)}$ in line~\ref{alg:MU:Phi} is the main focus of the results presented here. 
\clearpage

\begin{algorithm}[ht!]
\begin{itemize}
	\item $\boldsymbol{\mathscr{X}}$ is a tensor of size $I_{1} \times I_{2} \times \dots I_{N}$; $\boldsymbol{\mathscr{M}} = [ \bm{\lambda}; \mathbf{A}^{(1)} \dots \mathbf{A}^{(N)} ]$ is a rank-$R$ initial guess
	\item $k_{\mbox{max}}$ and $\ell_{\mbox{max}}$ are the maximum outer and inner iterations, respectively
\end{itemize}
\begin{algorithmic}[1]
	\FOR {$k=1, \dots, k_{\mbox{max}}$}
	    \FOR {$n=1, \dots, N$}
	    \STATE $ \mathbf{B}^{(n)} \gets (\mathbf{A}^{(n)} +\mathbf{S}) \bm\Lambda$ \qquad\COMMENT{$\triangleright$ \textit{$\mathbf{S}$ is used to remove inadmissible zeros}}
	    \STATE $\bm\Pi^{(n)} \gets (\mathbf{A}^{(N)} \odot \dots \odot \mathbf{A}^{(n+1)} \odot \mathbf{A}^{(n-1)} \odot \dots \mathbf{A}^{(1)} )^{T} $ 
	        \FOR {$\ell=1, \dots,  \ell_{\mbox{max}}$}
	            \STATE $\bm\Phi^{(n)} \gets (\mathbf{X}_{(n)} \oslash \max(\mathbf{B}^{(n)} \bm\Pi^{(n)}, \epsilon) ) (\bm\Pi^{(n)})^{T}$
	            \qquad\COMMENT{$\triangleright$ \textit{$\oslash$ denotes elementwise division}}\label{alg:MU:Phi}
	            \STATE $ \mathbf{B}^{(n)} \gets \mathbf{B}^{(n)} * \bm\Phi^{(n)}$
	            \qquad\COMMENT{$\triangleright$ \textit{$*$ denotes elementwise multiplication}}
	        \ENDFOR
	        \STATE $\bm\lambda = \mathbf{e}^{T} \mathbf{B}^{(n)}$
	        \STATE $\mathbf{A}^{(n)} \gets \mathbf{B}^{(n)}\bm\Lambda^{-1} \text{, where } \bm{\Lambda} = \text{diag}(\bm\lambda)$
	    \ENDFOR
	\ENDFOR
\end{algorithmic}
\caption{CP-APR MU (adapted from Algorithm 3 in~\cite{doi:10.1137/110859063})}
\label{alg:MU}
\end{algorithm}

%
\paragraph*{Contributions}
We make two key contributions towards understanding the performance characteristics of the SparTen CP-APR MU implementation:
\begin{enumerate}
    \item \emph{Performance analysis} of the $\bm\Phi^{(n)}$ matrix computation kernel, which dominates the overall execution time of the CP-APR MU algorithm.
    We demonstrate that the $\bm\Phi^{(n)}$ computation is memory-bound, and that two suspected performance bottlenecks---atomic operations and indirect pointer access---have significant impact on performance.
    \item \emph{Grid search results} for $\bm\Phi^{(n)}$ matrix computation on the Kokkos parallel policy on CPU and GPU architectures.
    Our analysis demonstrates that better policy parameter selection can provide an average of $2.25\times$ and $1.7\times$ speedup on CPUs and GPUs, respectively.
\end{enumerate}
\section{Background}
\label{sec:background}
The Canonical Polyadic Alternating Poisson Regression (CP-APR) algorithm generates a non-negative approximation of a tensor with Poisson distribution (i.e., count data).
There are three primary methods for calculating CP-APR:
\begin{enumerate*}[label=(\roman*)]
    \item MU,
    \item Projected Damped Newton for the Row Subproblem (PDNR), and
    \item Projected Quasi-Newton for the Row Subproblem (PQNR)
\end{enumerate*}.
PDNR and PQNR require fewer iterations to converge due to their use of Newton's method to solve the problem at the granularity of rows (i.e., each row can converge to its solution independently to other rows)~\cite{HaPlKo15}.
However, MU provides a more straight-forward parallel implementation based on dense matrix operations, and as a result, yields better overall performance on parallel systems.
Therefore, as a first-step, we focus our efforts on the MU algorithm for CP-APR.

Pressure Point Analysis (PPA)~\cite{czechowski2019diagnosing} entails systematically testing \emph{pressure points}, which are hardware resources hypothesized (but not necessarily known) to be performance bottlenecks. 
PPA temporarily disregards program correctness and ``perturbs'' the code by changing the \emph{utilization} of the targeted hardware resource.
If there is a substantial performance increase or decrease, then the targeted hardware resource is likely a true bottleneck, and we can begin efforts on addressing it.
PPA assists in code optimization by determining the upper bound on expected performance with minimal changes to the code.
An example of PPA in evaluating cache as a bottleneck is to reduce access to a large matrix to just a single row.
This limits the access to the L1 cache, and if performance increases significantly, we know that better data reuse will likely lead to better performance.


The Roofline Model~\cite{10.1145/1498765.1498785} is a simple performance model for determining the upper bound on performance for a given algorithm, expressed by its \emph{operational intensity}, on a given system.
Operational intensity of an algorithm is the ratio between work $W$ (e.g., floating point operations) and data access $Q$ (e.g., bytes), as shown in Equation~\ref{eq:ai}.
\begin{equation}
    I = \frac{W}{Q}
    \label{eq:ai}
\end{equation}

A given hardware system is represented in the model as as a plateaued line, representing the equation 
\begin{equation}
    P = \min{\left(\pi, \beta I\right)} \; ,
    \label{eq:rm}
\end{equation}
where $P$ is the \emph{attainable} compute performance, $\pi$ is the \emph{peak} compute performance, and $\beta$ is the memory bandwidth of the system.
The point at which Equation~\ref{eq:rm} plateaus is known as the system's \emph{balance} point.
If the operational intensity of an algorithm is on the left side of the system's balance on the x-axis (i.e., operational intensity value is low), the algorithm will be memory bandwidth-bound (or simply memory-bound);
otherwise, it will be compute-bound (i.e., capable of achieving peak performance on the system).
Figure~\ref{fig_roofline_cpu} shows the Roofline Model for the $\bm\Phi^{(n)}$ matrix computational kernel (vertical green bar at $I = 0.25$) on the Intel E5-2690v4 CPU (blue line).

We aim to first use the Roofline Model to determine whether key compute kernels of CP-APR MU are memory-bound or compute-bound, and to what extent, and then use the PPA to probe which specific hardware resources limit their performance.
\begin{figure}[!htb]
	\centering
	\includegraphics[width=0.625\textwidth]{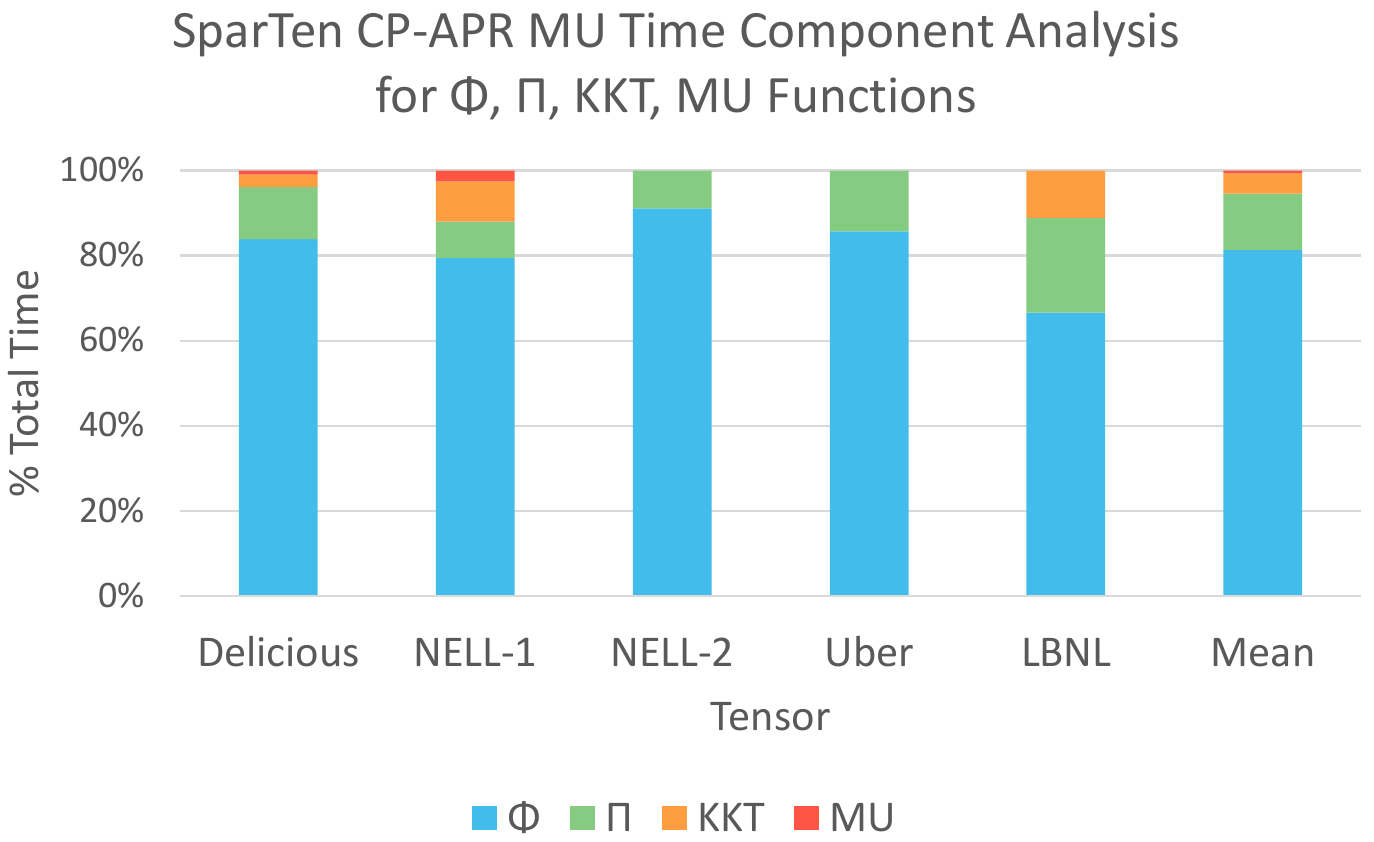}
	\caption{Runtime analysis for SparTen CP-APR MU kernels. The four kernels are for computing $\bm\Phi^{(n)}$, $\bm\Pi^{(n)}$, KKT conditions, and MU matrix product updates.}
	\label{fig_region_runtime}
\end{figure}

\section{Methods}
\label{sec:methods}


We omit a detailed discussion of the CP-APR MU algorithm for brevity, and direct interested readers to~\cite{doi:10.1137/110859063}.
At a high level, it is an algorithm that iteratively computes a solution (i.e., factor matrices) using a series of matrix operations. In the following sections, we present the parallel implementation details, roofline model analysis, and pressure point analysis of the CP-APR MU implementation in SparTen. 

\subsection{CP-APR MU Implementation and Parallelization}
\label{subsec:cpaprmu}
We begin our analysis of CP-APR MU by employing the \emph{SimpleKernelTimer} profiling routine from the Kokkos Profiling and Debugging Tools to identify compute kernels that dominate the overall execution time of the SparTen implementation.
Figure~\ref{fig_region_runtime} shows a breakdown of the execution time for the four most time consuming compute kernels---calculating $\bm\Phi^{(n)}$, $\bm\Pi^{(n)}$, KKT conditions, and MU matrix product updates---on five representative tensors from the FROSTT benchmark dataset~\cite{frosttdataset}.
As seen in the figure, the computation of the $\bm\Phi^{(n)}$ matrix comprises, on average, $81\%$ of the execution time among these four kernels.
As such, we will focus on analyzing and optimizing the $\bm\Phi^{(n)}$ matrix computation kernel, and refer to it simply as the $\bm\Phi^{(n)}$ kernel.
The $\bm\Phi^{(n)}$ kernel calculation, where $n$ is the mode index of the outer loop of the CP-APR MU algorithm, is shown in Algorithm~\ref{alg:cpaprmu}.

\begin{algorithm}[!htb]
\begin{itemize}
	\item $\boldsymbol{\mathscr{X}}$ is a tensor of size $I_{1}\times I_{2}\times\cdots I_{N}$ ($\mathbf X _ {(n)}$ is the mode-$n$ matricization of $\boldsymbol{\mathscr{X}}$)
	\item $R$ is the desired number of components (i.e., rank) in the model
	\item $\mathbf{B}$ is an $I _ n \times R$ dense matrix
	\item $\bm\Pi$ is an $R \times J _ n$ dense matrix, where $J _ n = \prod _ {m \neq n} I _ m$
	\item $\epsilon$ is the minimum divisor value to prevent divide-by-zero numerical issues
\end{itemize}
\begin{algorithmic}[1]
	\STATE $\bm\Phi^{(n)} \gets
	\left(
	\mathbf{X} _ {(n)} \oslash \max \left(
	\mathbf{B} \, \bm\Pi , \epsilon
	\right)
	\right)  \bm\Pi ^ \top$ 
	\qquad\COMMENT{$\triangleright$ \textit{$\oslash$ is elementwise division, $\max$ operates elementwise}}
\end{algorithmic}
	\caption{CP-APR MU $\bm\Phi^{(n)}$ calculation}
	\label{alg:cpaprmu}
\end{algorithm}
%
%

One important point to note is that because CP-APR is intended for use with \emph{sparse} tensors, forming the $\mathbf{X}_{(n)}$ and $\bm\Pi$ matrices explicitly is unnecessary, particularly because these are extremely large matrices.
For example, for a 4-way tensor of size $1,000\times 1,000\times 1,000\times 1,000$, a CP-APR MU decomposition with rank $R = 10$ will yield matrices $\mathbf{X}_{(n)}\in\mathbb{R}^{1,000\times 10^9}$ and $\bm\Pi\in\mathbb{R}^{10\times 10^9}$.
Instead, most high-performance implementations calculate the result one non-zero element at time, which greatly reduces the size of the intermediate data structures and thereby greatly improves the overall performance.
However, this method of calculating the result one non-zero element at a time leads to a race condition in a multi-threaded implementation.
If two non-zero elements update the same row in $\bm\Phi^{(n)}$, the updates must be serialized to maintain correctness.

There are two strategies for addressing this race condition.
The first strategy is simply to use atomic operations during updates to $\bm\Phi^{(n)}$ to ensure serialization.
The second strategy involves sorting the non-zero elements such that non-zero elements that update the same row in $\bm\Phi^{(n)}$ are stored contiguously, as introduced by Phipps and Kolda for a related tensor decomposition method~\cite{doi:10.1137/18M1210691}.
When assigning contiguous non-zero elements to threads, this strategy will maximize the likelihood that non-zero elements that update the same row are assigned to the same thread.
Then, atomic operations are required only at non-zero element ``boundaries'' (i.e., where non-zero elements go from updating row $i$ to row $i+1$) to ensure correctness.
To reduce the number of sorting operations from once every mode within every iteration, the sorting can be done for each mode at the very beginning, and the sorting information can be stored in $N$ permutation arrays, one for each mode.

The impact of each strategy on performance depends heavily on the target architecture.
Atomic operations scale poorly with a large number of threads, as having more threads increases the probability of contention for updates between threads.
Using a permutation array leads to indirect and scattered memory access, which performs poorly on modern memory systems.
As a result, SparTen utilizes slightly different implementations on CPUs and GPUs to maximize performance.
While we acknowledge that this goes against the idea of a \emph{single} performance portable implementation, we note that this is nevertheless much more portable than using two different programming models (e.g., OpenMP/CilkPlus/TBB on CPUs and CUDA on GPUs). This strategy effectively provides a composite implementation which targets separately two broad classes of processors, namely multi-core CPUs and highly parallel GPUs, where the desired implementation is chosen at compile time.

Algorithms~\ref{alg:phi-gpu} and~\ref{alg:phi-cpu} show the SparTen implementations of the $\bm\Phi^{(n)}$ kernel for GPUs and CPUs, respectively.
The GPU implementation is the simpler of the two, with each thread assigned to one non-zero element and using atomic operations to update $\bm\Phi^{(n)}$ (line~\ref{eq_atomic_gpu}).
The CPU implementation uses a similar algorithm but adds an \emph{atomic mitigation} method.
Because the non-zero elements are sorted (via the permutation array $\mathbf{P}$, line~\ref{eq_indirect_cpu}), each thread falls under one of three cases:
\begin{enumerate}
    \item Both current thread and previous thread(s) have non-zero elements with the same coordinate $i$.
    \item Current thread has \emph{every} non-zero element with coordinate $i$.
    \item Both current and next thread(s) have non-zero elements with the same coordinate $i$.
\end{enumerate}
In each case, the results are accumulated to a local array and then written out to $\bm\Phi^{(n)}$ when the coordinate value changes.
For case 2), since every non-zero element that updates row $i$ belongs to the current thread, writing the accumulated result to $\bm\Phi^{(n)}$ does \emph{not} require an atomic operation.
On the other hand, cases 1) and 3) require atomic operations to update $\bm\Phi^{(n)}$ as other threads may be updating the same row in $\bm\Phi^{(n)}$.

\begin{algorithm}[!htb]
	\begin{itemize}
		\item $nnz$ is the total number of non-zero elements.
		\item $\mathbf{I}[n]$ is the permutation array for mode $n$.
	\end{itemize}
	\begin{algorithmic}[1]
		\FOR {$j = 0$, \dots, $nnz-1$}
		\STATE $i \gets \mathbf{I} [n][j]$ \qquad\COMMENT{$\triangleright$ \textit{coordinate array}} \label{alg:line:perm2}
		\STATE $s \gets 0$
		\FOR {$r =0, \dots, R-1$}
		\STATE $s \gets s + \mathbf{B} [i][r] * \bm\Pi [j][r]$\label{alg:phi_gpu_inner_product}
		\ENDFOR
		\STATE $s \gets \mathbf{X}_{(n)} [j] / \max (s, \epsilon)$
		\FOR {$r =0, \dots, R-1$}
		\STATE $\bm\Phi [i][r] \gets \bm\Phi [i][r] + s * \bm\Pi  [j][r]$ \qquad\COMMENT{$\triangleright$ \textit{atomic update}}  \label{eq_atomic_gpu}
		\ENDFOR
		\ENDFOR
	\end{algorithmic}
	\caption{CP-APR MU $\bm\Phi^{(n)}$ calculation (GPU). Each thread assigned $1$ non-zero element.}
	\label{alg:phi-gpu}
\end{algorithm}

\begin{algorithm}[!htb]
	\begin{algorithmic}[1]
		\FOR {$k = 0, \dots, nthreads-1$}
		\FOR {$z = kV, \dots, kV + V -1$}
		\IF {$z \geq$ nnz}
		\STATE $\textrm{continue}$
		\ENDIF
		\STATE $j \gets P[n][z]$ \label{eq_indirect_cpu} \qquad\COMMENT{$\triangleright$ \textit{permutation array}} \label{alg:line:perm}
		\STATE $i \gets I [n] [j]$
		\STATE $s \gets 0$
		\FOR {$r = 0, \dots, R-1$}
		\STATE $s \gets s + B [i] [r] * \Pi [j] [r]$
		\ENDFOR
		\STATE $s \gets X [j] / \max (s, \epsilon)$
		\IF{thread $k$ has every non-zero with coordinate $i$}
		\FOR {$r = 0, \dots, R-1$}
		\STATE $\Phi [i][r] \gets \Phi [i][r] + \textrm {tmp} [r]$ \qquad\COMMENT{$\triangleright$ \textit{non-atomic}} \label {eq_nonatomic_cpu} 
		\ENDFOR
		\ELSE
		\FOR {$r = 0, \dots, R-1$}
		\STATE $\Phi [i][r] \gets \Phi [i][r] + \textrm {tmp} [r]$ \qquad\COMMENT{$\triangleright$ \textit{atomic}} \label {eq_atomic_cpu}
		\ENDFOR
		\ENDIF
		\ENDFOR
		\ENDFOR
	\end{algorithmic}
	\caption{CP-APR MU $\bm\Phi^{(n)}$ calculation (CPU). Each thread assigned $V$ non-zeros elements.}
    \label{alg:phi-cpu}
\end{algorithm}

\subsection{Roofline Model Analysis}
\label{subsec:roofline}
With parallel implementations now available in SparTen, we analyze the baseline performance of the CP-APR MU $\bm\Phi^{(n)}$ computation using the Roofline Model. 
We analyze the baseline algorithm, shown in Algorithm~\ref{alg:cpaprmu}, to determine the number of floating point operations and data accesses, and calculate the operational intensity, as shown in Equations~\ref{eq:w}--\ref{eq:oi}.
\emph{Words} are 8 bytes (i.e., double-precision) in this example.

\begin{align}
    \label{eq:w} 
    W & = \textrm{nnz}(4R + 2) & \textrm{FLOPs} \\
    Q & = \textrm{nnz}(5R + 2) & \textrm{Words}\\
    I & = \frac W Q = 0.125 & \textrm{Operational intensity}\label{eq:oi}
\end{align}

Operational intensity of $0.125$ is extremely low, making it likely memory-bound on most modern systems.
We can further refine this specifically for the CPU implementation, as it uses the atomic mitigation technique.
For the CPU implementation, we have

\begin{align}
    W & = \textrm{nnz}(4R + \frac R V + 3)  & \textrm{FLOPs}\\
    Q & = \textrm{nnz}(6R + \frac {2R} V + 3) & \textrm{Words}\\
    I & = \frac W Q \approx 0.27 & \textrm{Operational intensity}
\end{align}

We also require the peak compute performance and the memory bandwidth of the target runtime system to generate the Roofline Model.
We can calculate the peak computer performance $\pi$, in GFLOP/s using the following equation for the Intel E5-2690v4 CPU:

\begin{equation}
\begin{split}
    \pi  = & \ \textrm {(clock speed)}
           \times \textrm {(core count)}
            \times \textrm {(operations per cycle)}
            \times \textrm {(processor count)} \notag \\
         =  & \  2.6 \times 14 \times 16 \times 2 \\
         =  & \  1164.8 \ \textrm{GFLOP/s}\notag \; .
\end{split}
\end{equation}

The same equation can be used to calculate the peak performance for the NVIDIA K80 GPU, which has a peak compute performance of approximately 2910 GFLOP/s.
As for memory bandwidth, we use vendor published numbers, which are 153.6 GB/s and 480 GB/s for Intel E5-2690v4 and NVIDIA K80, respectively.

\begin{figure}[!htbp]
\centering
\includegraphics[width=0.8\textwidth]{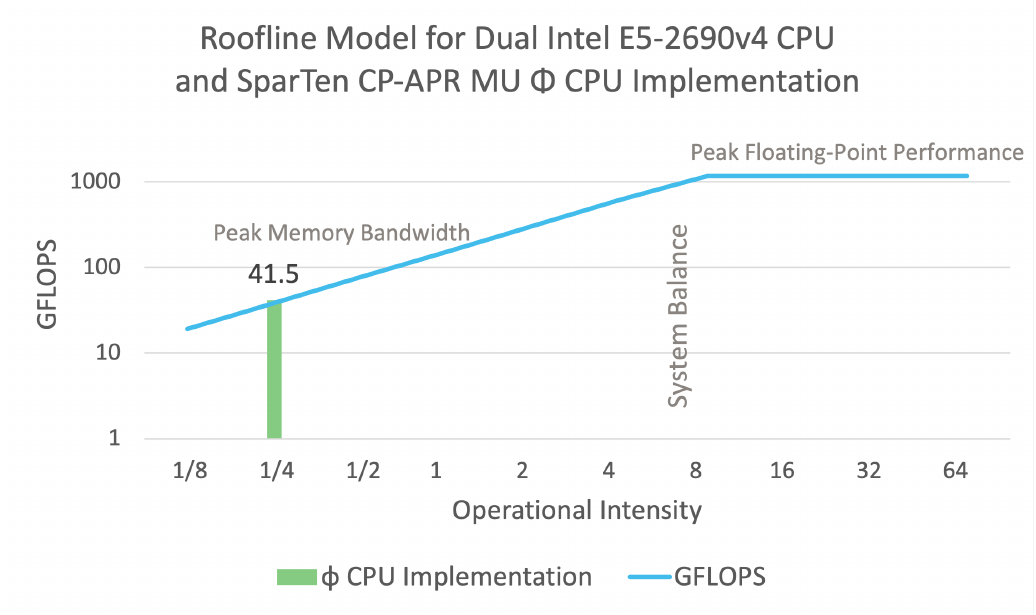}
\caption{Roofline model for $\bm\Phi^{(n)}$ CPU implementation on dual Intel E5-2690v4.}
\label{fig_roofline_cpu}
\end{figure}

\begin{figure}[!htbp]
\centering
\includegraphics[width=0.8\textwidth]{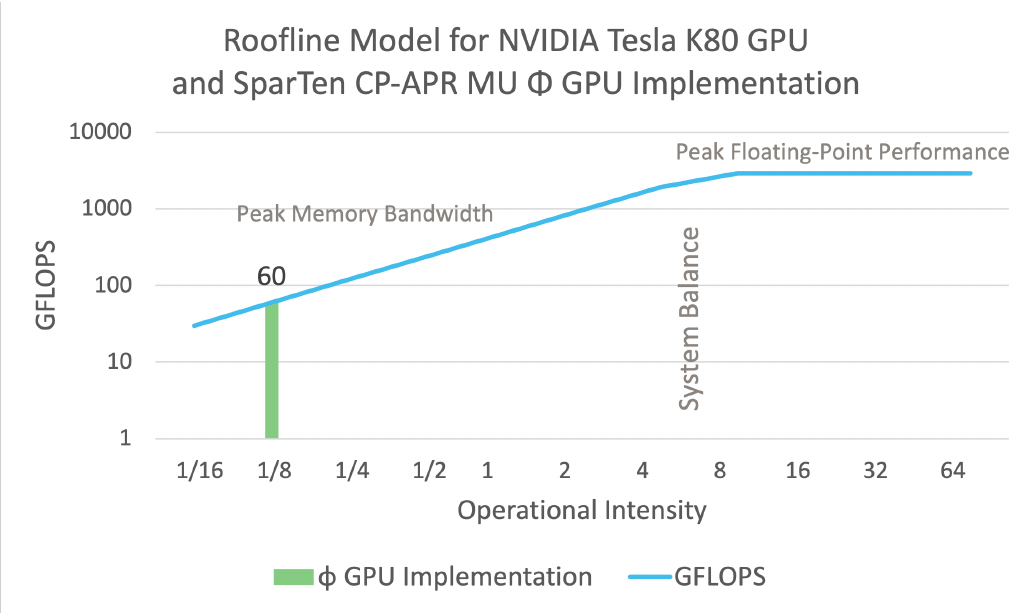}
\caption{Roofline model for $\bm\Phi^{(n)}$ GPU implementation on NVIDIA Tesla K80.}
\label{fig_roofline_gpu}
\end{figure}

We can now generate the Roofline Model for the $\bm\Phi^{(n)}$ kernel on these two systems. Figures~\ref{fig_roofline_cpu} and~\ref{fig_roofline_gpu} illustrate the $\bm\Phi^{(n)}$ computation is severely memory-bound for both CPUs and GPUs, respectively.
The expected performance is $41.5$ GFLOP/s for the CPU and $60$ GFLOP/s for the GPU, which are only small fractions of their peak compute performance values.
This suggest that we should focus our efforts on optimizing data access (e.g., via better caching, operation fusion to minimize intermediate data, etc.)

\subsection{Pressure Point Analysis}

PPA provides a more systematic approach to identifying performance bottlenecks and their impact on overall performance of an algorithm.
From the Roofline Model, we determined that the $\bm\Phi^{(n)}$ kernel is \emph{primarily} limited by the time to access data from the main memory.
The next step is to determine exactly which specific data accesses are limited and which hardware resources are causing it.

\subsubsection{Atomic operations}
On both the CPU and GPU implementations of the $\bm\Phi^{(n)}$ kernel, atomic operations are used to update the $\bm\Phi^{(n)}$ matrix rows to ensure serialized updates between different threads that are working on the same row at the same time.
Atomic add (i.e., \emph{atomic\_add(a,b)} for $a = a + b$) is a major source of memory reads and writes, and particularly expensive on modern parallel processors. As such, analysis of the use of atomic adds in CP-APR MU is a good candidate for PPA.
To run PPA, we change the algorithm by replacing atomic operations with non-atomic operations (i.e., $a = a + b$).
Specifically, line~\ref{eq_atomic_gpu} from Algorithm~\ref{alg:phi-gpu} (GPU) and line~\ref{eq_atomic_cpu} from Algorithm~\ref{alg:phi-cpu} (CPU) are replaced for this analysis.



\subsubsection{Data reuse}
For algorithms that are limited by data access, higher cache hit rate will result in better performance.
PPA can also be used to determine how the performance will change when data are accessed from the cache more frequently.
We change the CPU and GPU algorithms by limiting the access to every matrix to a small subset of rows (e.g., for a $1000\times 1000$ matrix, we hard-code the implementation to access the first matrix row only, regardless of what row is supposed to be accessed).
We applied this strategy to different combinations/permutations of matrices involved in the $\bm\Phi^{(n)}$ computation, but saw small improvements in performance for each individual change.
We observed significant improvement when we applied this change for every matrix, which simulates a perfect data reuse in cache, and therefore, provides an upper bound on the achievable performance for the kernel.
The result of our PPA is described, with analysis, in Section~\ref{sec:exp}.
\section{Experimental Results}
\label{sec:exp}
We describe our experimental setup and data, and present several key results from our experiments.

We conduct our PPA and policy study experiments using the CPU and GPU system, shown in Table~\ref{tab:system}, on six real-world sparse tensors from the FROSTT tensor benchmark dataset~\cite{frosttdataset}, shown in Table~\ref{tab:tensors}.
Our timing results reflect averages of five runs per experiment.

\begin{table}[!htbp]
\centering
    \caption{Experimental setup}
    \begin{tabular}{c|c|c} 
         Type & Name & \# Cores \\ 
         \hline
         \hline
         CPU & Intel E5-2690v4 & 14 $\times$ 2 (dual-socket)\\ 
         \hline
         GPU & NVIDIA Tesla K80 & 4992 (CUDA cores)\\
         \hline
    \end{tabular}
\label{tab:system}
\end{table}

\begin{table}[!htbp]
\begin{center}
\caption{Tensors used for evaluation}
\begin{tabular}{l|l|c} 
 Tensor & Dimensions & Number of Non-zero Elements \\ 
 \hline\hline
 Chicago-Crime & 6.2K $\times$ 24 $\times$ 77 $\times$ 32 & 5.3M \\ 
 Enron & 6.1K $\times$ 5.7K $\times$ 244K $\times$ 1.2K & 54M \\ 
 LBNL-Network & 1.6K $\times$ 4.2K $\times$ 1.6K $\times$ 4.2K $\times$ 868K & 1.7M \\
 NELL-2 & 12.1K $\times$ 9.2K $\times$ 28.8K & 76.9M \\ 
 NIPS & 2.5K $\times$ 2.9K $\times$ 14.0K $\times$ 17 & 3.1M \\
 Uber & 183 $\times$ 24 $\times$ 1.1K $\times$ 1.7K & 3.3M \\
\hline
\end{tabular}
\label{tab:tensors}
\end{center}
\end{table}

\subsection{Experiment 1: Pressure Point Analysis}

After identifying the $\bm\Phi^{(n)}$ kernel within and CP-APR MU as the most time consuming kernel (Section~\ref{subsec:cpaprmu}) and identifying the kernel as memory-bound (Section~\ref{subsec:roofline}), we introduce PPA on the atomic operations by modifying the SparTen CP-APR MU $\bm\Phi^{(n)}$  implementation.
Specifically, we modify the atomic operations in line~\ref{eq_atomic_gpu} from Algorithm~\ref{alg:phi-gpu} (GPU) and line~\ref{eq_atomic_cpu} from Algorithm~\ref{alg:phi-cpu} (CPU).

For analyzing the impact of higher cache reuse and a more regular memory access pattern (i.e., not using a permutation array, which leads to scattered memory access), we limit memory access by having each thread access only a particular row within the matrices involved in the $\bm\Phi^{(n)}$ calculation.
While the impact of perturbing the access to only a \emph{subset} of the matrices involved in the $\bm\Phi^{(n)}$ calculation was small, perturbing the access to every matrix showed non-trivial improvement in performance.
Finally, we combine the perturbation for both atomic operations and data reuse to demonstrate an upper bound on the achievable performance---i.e., if no thread contention and ``perfect'' memory access can be achieved.

\subsubsection{PPA Results on a CPU}
\begin{figure}[!htbp]
\centering
\includegraphics[width=0.8\textwidth]{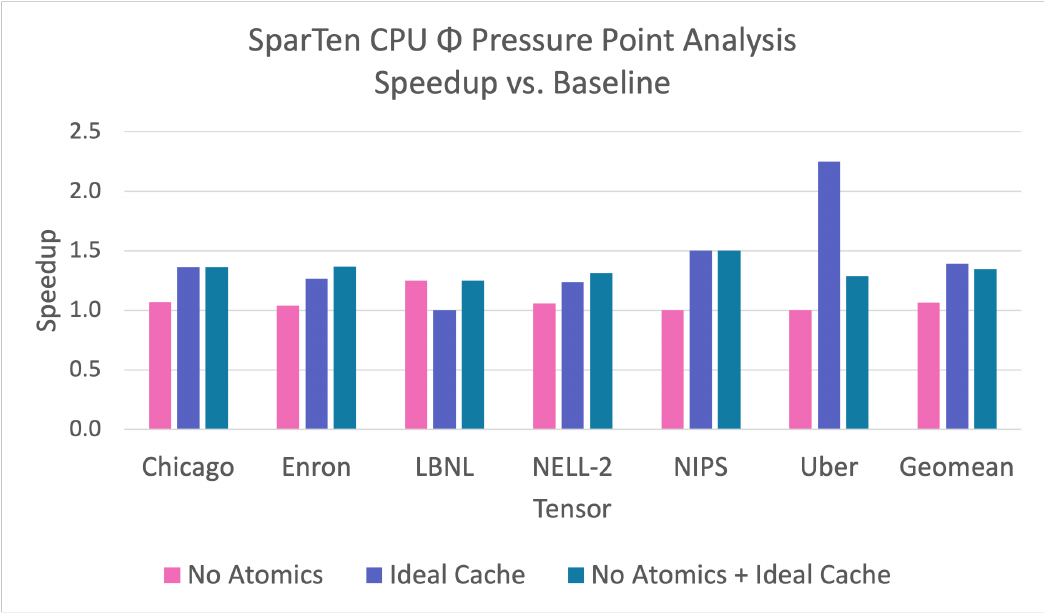}
\caption{PPA result for SparTen CP-APR MU $\bm\Phi^{(n)}$ computation on Intel Xeon E5-2690v4 CPU.}
\label{fig_cpu_ppa}
\end{figure}

Figure~\ref{fig_cpu_ppa} shows the result of our PPA on the CPU, where timing speedups of computing $\bm\Phi^{(n)}$ over a baseline (vertical axis) are plotted for each of the data tensors along with the geometric mean ({\it geomean}) of the speedups (horizontal axis).
The speedup for using no atomic operations over the baseline (i.e., using atomic operations) ranges from $1.0\times$ to $1.3\times$ (magenta bars), with a geometric mean speedup of $1.1\times$.
The speedup from perfect data reuse in cache and regular memory access ranges from $1.0\times$ to $2.3\times$ (blue bars), with a geometric mean speedup of $1.4\times$.
Finally, when both PPA perturbations are combined, we see a speedup that ranges from $1.3\times$ to $1.5\times$ (teal bars).
We see that, in most cases, there is an additive effect in combining both perturbations, with results for the \emph{Uber} tensor being the only exception.

Results for the \emph{Uber} tensor are counter intuitive, as the small mode lengths should, in theory, yield higher speedups by eliminating the need for atomic operations, as the fewer number of rows accessed across the threads should decrease the probability of contention.
Additionally, the small sizes of the factor matrices and the fewer number of non-zero elements should allow the tensor to have high cache reuse even without the PPA perturbation; therefore we expected little performance improvement from our PPA data reuse perturbation.
However, \emph{Uber} demonstrated the highest speedup from the data reuse perturbation among our six evaluation tensors.
Our hypothesis is that the non-zero element \emph{sparsity pattern} within the tensor creates skewed memory accesses, leading to this counter intuitive result.

\subsubsection{PPA Results on a GPU}
\begin{figure}[!htbp]
\centering
\includegraphics[width=0.8\textwidth]{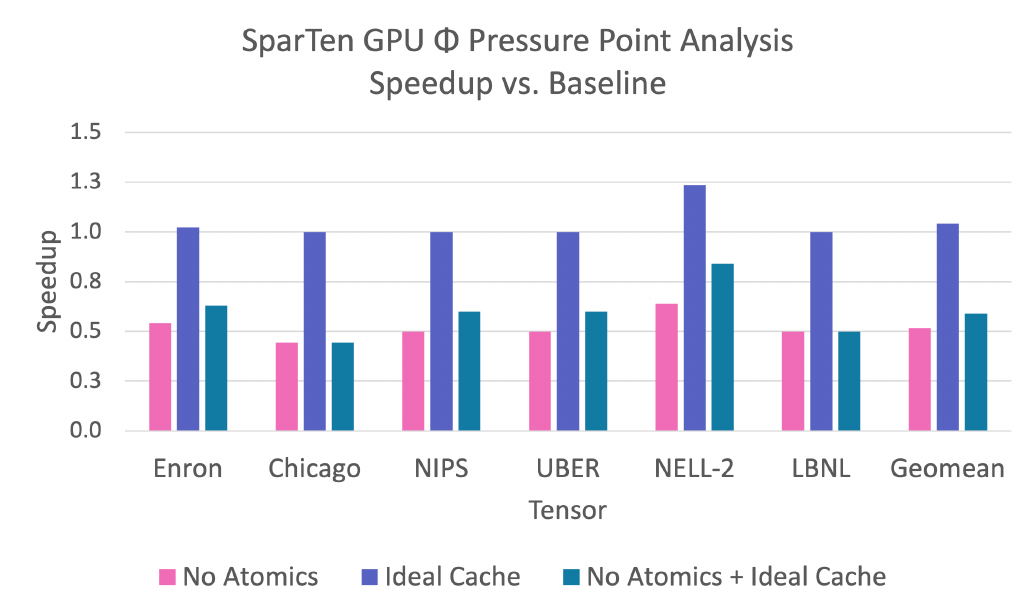}
\caption{PPA result for SparTen CP-APR MU $\bm\Phi^{(n)}$ computation on NVIDIA Tesla K80 GPU.}
\label{fig_gpu_ppa}
\end{figure}

Figure~\ref{fig_gpu_ppa} shows the corresponding results of our PPA on the GPU. On the GPU, preventing the use of atomic operations actually \emph{hurts} performance, and we see a slowdown ranging from $0.44\times$ to $0.66\times$.
This is likely caused by the GPU's architectural feature that forces atomic operations to go through the L2 cache, thereby avoiding portions of the memory hierarchy entirely compared with non-atomic operations. This feature 
is likely a design decision due to the inherent difficulty in implementing atomic operations with \emph{tens of thousands} of concurrent threads on GPUs.
The speedup from our data reuse perturbation ranges from $1.0\times$ to $1.2\times$.
While this is low, it is not entirely surprising given the small cache sizes on GPUs.
When we combine both perturbations, we see overall slowdowns ranging from $0.4\times$ to $0.8\times$.
This is due to the slowdown from using atomic overshadowing the already small benefit we see from perfect data reuse.


\subsection{Experiment 2: Performance of GPU Algorithm on CPU}
\label{subsec:gpu-cpu}
In seeking to understand the characteristic differences between runtime performance of algorithms on CPUs and GPUs while exploring the performance portability potential of Kokkos, we modify the GPU implementation to run on CPU by using the same Kokkos parallel policy as used in the SparTen CPU baseline (primarily, the same number of threads, which is lower on CPU). 
The benefit of such an approach is that one processor-agnostic implementation could be used for both CPU and GPU.
Moreover, the GPU implementation leverages Kokkos for more control over parallelization compared to the current CPU implementation. 
We compare the performance of running the GPU implementation on the CPU with the original CPU baseline implementation, including results from the atomic operation and data reuse PPA perturbations.

\begin{figure}[!hbt]
\centering
\includegraphics[width=0.8\textwidth]{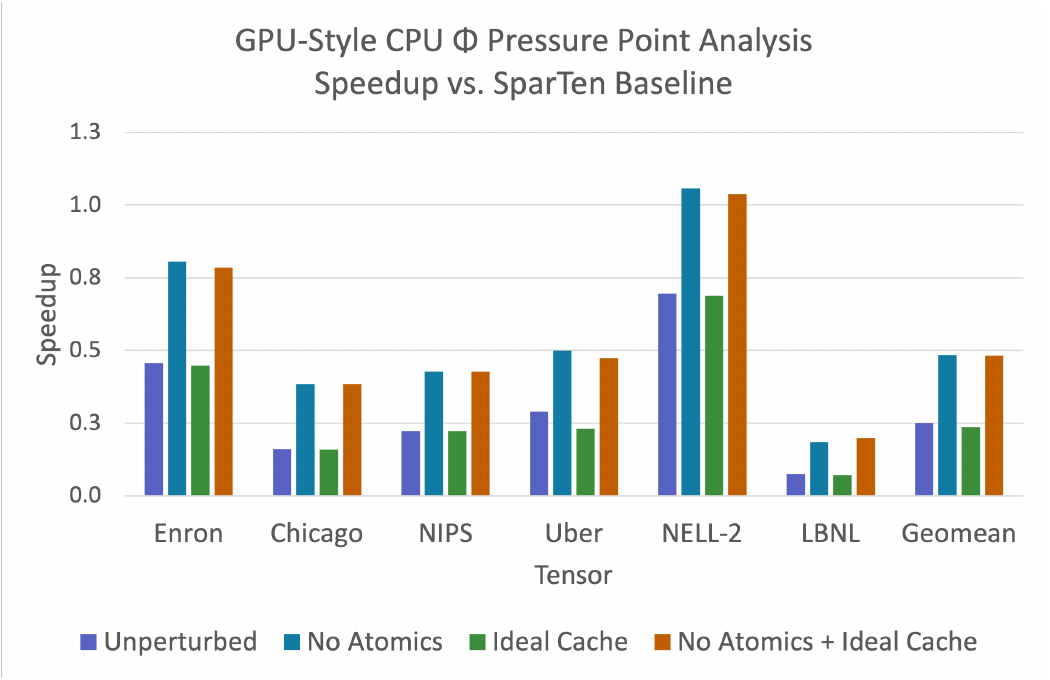}
\caption{Pressure point analysis results for GPU-style SparTen CP-APR MU $\bm\Phi^{(n)}$ computation on CPU. Unperturbed refers to the GPU-style implementation on CPU with no pressure point analysis perturbations, and all speedup is compared to the unperturbed SparTen CPU implementation.}
\label{fig_gpu_on_cpu}
\end{figure}

The results of this experiment are shown in Figure~\ref{fig_gpu_on_cpu}.
Compared with the original SparTen baseline, the GPU implementation on the CPU (Unperturbed) exhibits a slowdown ranging from $0.08\times$ to $0.7\times$ on the six input tensors. 
With the atomic operation perturbation (teal bars), GPU-style implementation on CPU shows speedup ranging from $0.2\times$ to $1.1\times$ over the baseline, and for the data reuse perturbation (green bars), the speedup ranges from $0.07\times$ to $0.7\times$. 
When both perturbations are combined (red bars), GPU-style implementation on CPU achieves speedup in the range of $0.2\times$ to $1.0\times$ compared to the SparTen baseline. 
This relatively low performance achieved by the GPU-style implementation on CPU suggests that the additional CPU-specific optimization (i.e., atomic mitigation mechanism described in Section~\ref{subsec:cpaprmu}) in the SparTen baseline CPU implementation is effective.
However, this also raises the question as to whether the Kokkos policy used in the SparTen CPU baseline is appropriate for use in the GPU implementation on the CPU due to the inherently different levels of parallelism available on CPUs and GPUs.

\subsection{Experiment 3: Parameterized Parallel Policy}
\label{subsec:parampparpol}
To determine the impact of Kokkos policy on the performance of the GPU implementation on the CPU, we first study the impact of the choice of Kokkos policy for the GPU implementation running on GPU.
For this experiment, we modify the SparTen driver and $\bm\Phi^{(n)}$ GPU implementation to accept Kokkos policy parameters: league size, team size, and vector size. 
League size loosely corresponds with the number of teams of workers (threads) in total, team size corresponds with the number of workers (threads) per team, and vector size corresponds with the amount of work assignable to each worker (thread). 
Kokkos uses these terms for defining the three-level hierarchical parallelism model as demonstrated in the triply-nested loops in the SparTen GPU and CPU $\bm\Phi^{(n)}$ implementations.
The outermost for-loop iterations are generally each assigned to a team, the mid-level for-loop iterations are each assigned to a worker (thread), and the innermost for-loop iterations are capable of being executed in parallel by a worker under appropriate algorithmic and hardware-supported circumstances. 
By exposing the Kokkos policy parameters, we can rapidly explore the policy space to gain an understanding of whether the default SparTen policies are effective on CPU and GPU systems. 
Team and vector sizes can also be set automatically by Kokkos, which presents an opportunity to observe to what extent Kokkos handles performance portability via default policy parameterization.

\begin{figure}[!hbt]
\centering
\includegraphics[width=0.6\textwidth]{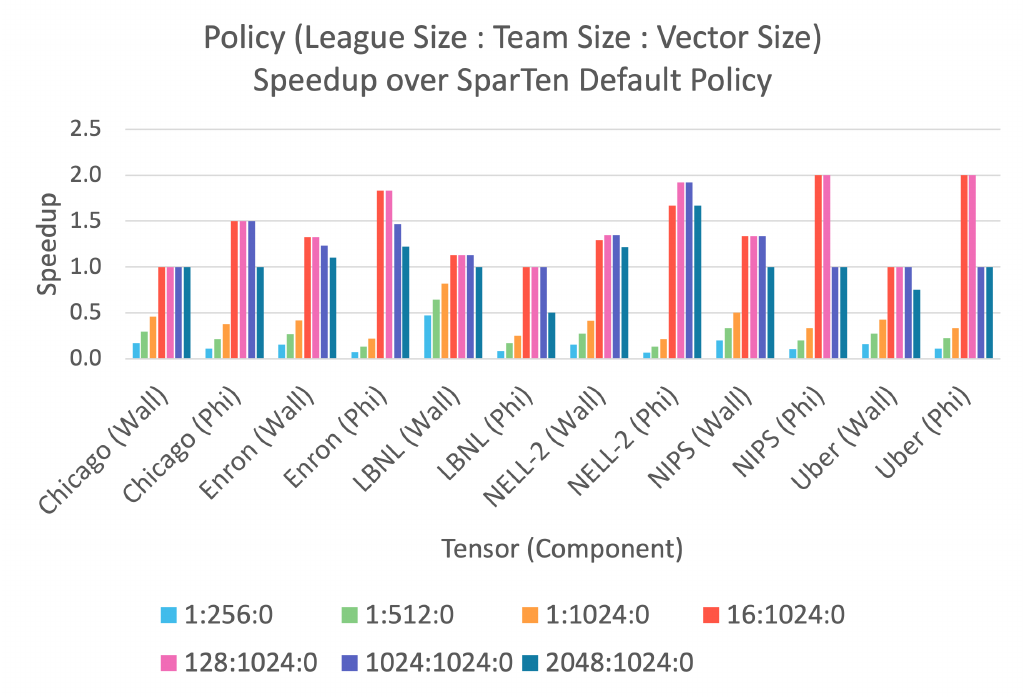}
\caption{Coarse parameter grid search for overall SparTen CP-APR MU and $\bm\Phi^{(n)}$ computation on GPU, varying league size and team size, with vector size unspecified and left to be chosen automatically.}
\label{fig_policy_initial}
\end{figure}

We compare the performance of the GPU implementation in terms of the entire SparTen CP-APR MU implementation (i.e., not just the $\bm\Phi^{(n)}$ calculation) and for just the $\bm\Phi^{(n)}$ computation with the unmodified implementation for seven policy configurations, varying the league and team size and leaving vector size to be determined by Kokkos. 
These results are shown in Figure~\ref{fig_policy_initial}, where {\it Wall} and {\it Phi} results correspond to those for the full CP-APR MU algorithm and just the $\bm\Phi^{(n)}$ computation, respectively.
For just the $\bm\Phi^{(n)}$ calculation, the geometric mean speedup over the six input tensors for the seven policy configurations yielded speedup ranging from $0.1\times$ to $1.7\times$, and for the entire CP-APR MU calculation, the geometric mean speedup over the six input tensors for the seven policy configuration ranged from $0.2\times$ to $1.2\times$. 
From this, we can conclude that Kokkos policy selection has a significant impact on GPU performance for the $\bm\Phi^{(n)}$ computation.
However, because the $\bm\Phi^{(n)}$ calculation speedup was significantly lower than that of the overall CP-APR MU computation, despite the $\bm\Phi^{(n)}$ computation making up $81\%$ of the four most time consuming kernels in CP-APR MU, the configuration may need to be adjusted dynamically for different parts of the algorithm to achieve maximum performance.

\subsection{Experiment 4: Kokkos Policy Study on GPU}
\begin{figure}[!hb]
\centering
\includegraphics[width=0.8\textwidth]{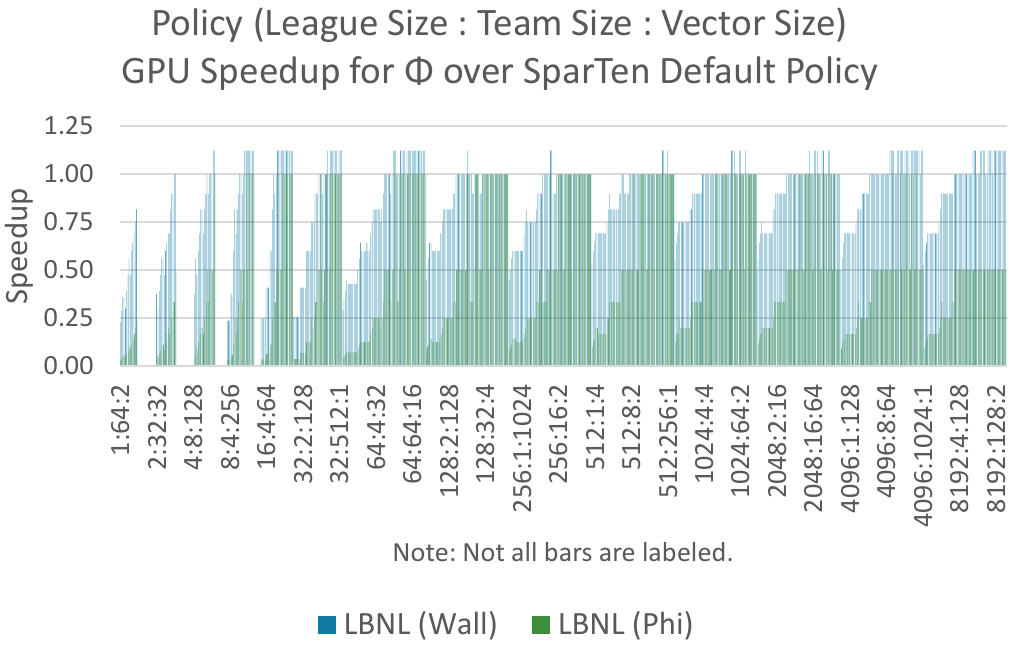}
\caption{Fine parallel policy parameter grid search for overall SparTen CP-APR MU and $\bm\Phi^{(n)}$ computation on GPU for {\it LBNL}, varying league size, team size, and vector size, for league sizes 1--8192. Not all bars are labeled.}
\label{fig_policy_gpu_lbnl_lower}
\end{figure}

\begin{figure}[!ht]
\centering
\includegraphics[width=0.8\textwidth]{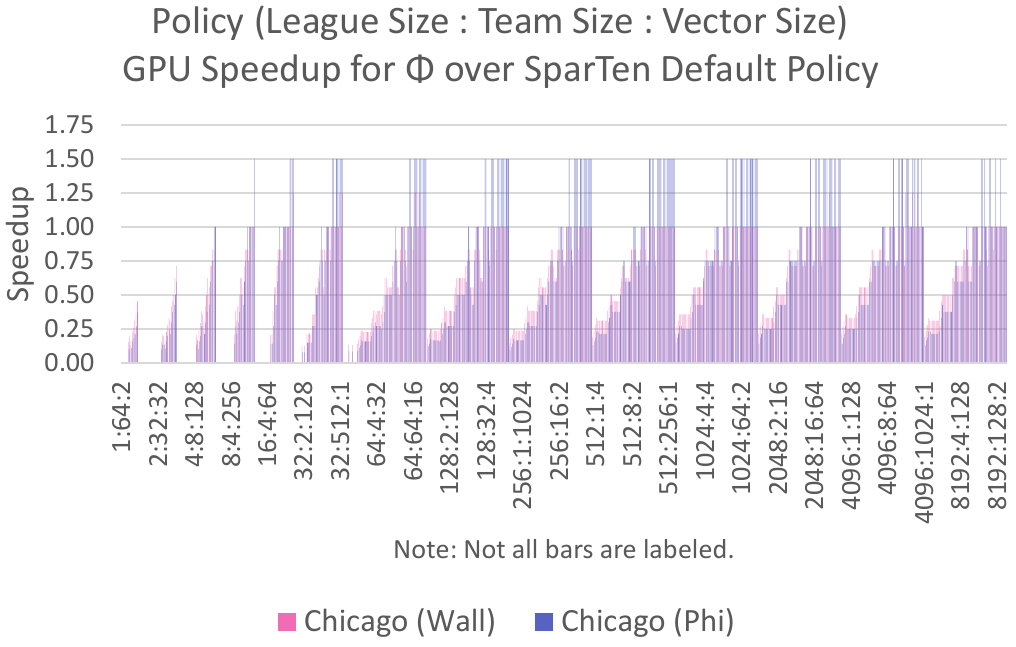}
\caption{Fine parallel policy parameter grid search for overall SparTen CP-APR MU and $\bm\Phi^{(n)}$ computation on GPU for {\it Chicago}, varying league size, team size, and vector size. Results for league sizes beyond 512 are not shown because they exhibit similar trends to league size 256. Not all bars are labeled.}
\label{fig_policy_gpu_chicago}
\end{figure}

We further extend our result from Section~\ref{subsec:parampparpol} by conducting an extensive grid-search over league size, using values from $1$ to $8192$, and for team size and vector size, using values from $1$ to combinations of these two sizes whose product equals $1024$.
The latter choice is due to a Kokkos requirement that the product between team size and vector size cannot exceed $1024$.
Due to the large search space, we perform this experiment using only the {\it LBNL} and {\it Chicago} tensors. Figure~\ref{fig_policy_gpu_lbnl_lower} and Figure~\ref{fig_policy_gpu_chicago} show the results for {\it LBNL} and {\it Chicago}, respectively.
For {\it LBNL}, speedups ranged from $0.0\times$ to $1.1\times$ for the overall CP-APR MU calculation and from $0.0\times$ to $1.0\times$ for the $\bm\Phi^{(n)}$ calculation.
This is in comparison to the result from Section~\ref{subsec:parampparpol}, where speedups ranged from $0.5\times$ to $1.1\times$ and $0.2\times$ to $1.0\times$ for CP-APR MU and $\bm\Phi^{(n)}$, respectively.
For {\it Chicago}, speedups ranged from $0.0\times$ to $1.3\times$ for the overall CP-APR MU calculation and from $0.0\times$ to $1.5\times$ for the $\bm\Phi^{(n)}$ calculation.
This is in comparison to the result from Section~\ref{subsec:parampparpol}, where speedup ranged from $0.2\times$ to $1.0\times$ and $0.1\times$ to $1.5\times$ for CP-APR MU and $\bm\Phi^{(n)}$, respectively.

Our more extensive parameter search suggests that a good heuristic could find the optimal policy and an exhaustive search is likely unnecessary.
Future study into finding a good heuristic for Kokkos policy will enable increased performance portability for CP-APR and potentially other Kokkos applications. 


\clearpage

\subsection{Experiment 5: Kokkos Policy Study of GPU Implementation on CPU}

Using the insight gained from the two previous experiments on the impact of Kokkos policy on GPUs, we now apply policy parameter grid search on the GPU implementation on CPU (from Section~\ref{subsec:gpu-cpu}). 
Initial exploration shows that unperturbed GPU implementation on CPU performs at best $0.7\times$ (from Figure~\ref{fig_gpu_on_cpu}) of the SparTen baseline CPU implementation.
However, we are interested in determining if this GPU implementation on CPU can further improve performance, when given the correct Kokkos policy, if the proper optimizations are applied (i.e., minimizing the impact of atomic operations and having better data reuse in cache).
As such, we apply our policy study on the perturbed implementation.

\begin{figure}[!t]
\centering
\includegraphics[width=0.8\textwidth]{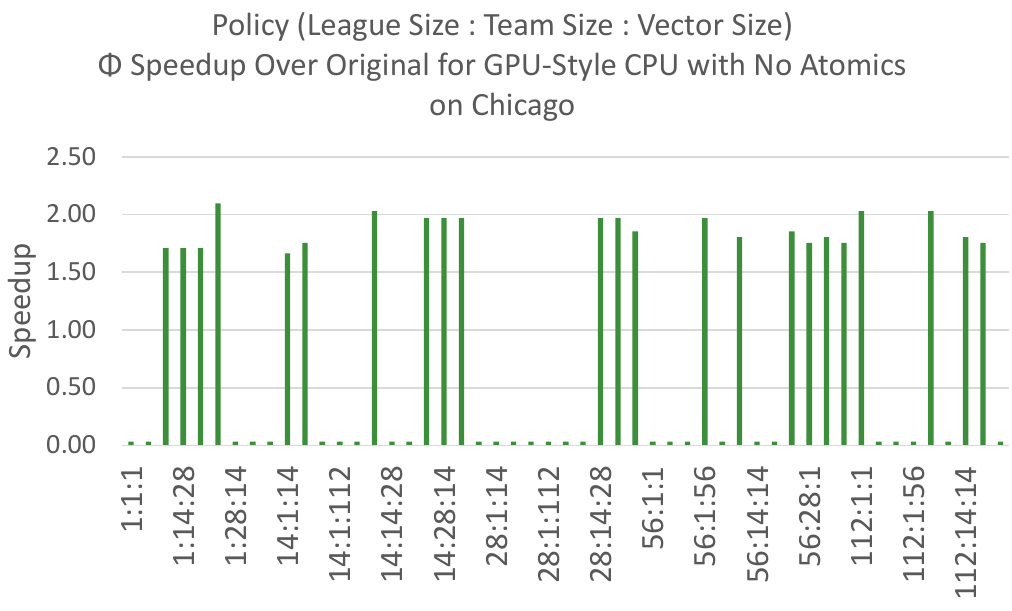}
\caption{Fine parallel policy parameter grid search for overall SparTen CP-APR MU and $\bm\Phi^{(n)}$  computation on CPU for {\it Chicago}, varying league size, team size, and vector size.}
\label{fig_policy_cpu_chicago}
\end{figure}

\begin{figure}[!t]
\centering
\includegraphics[width=0.8\textwidth]{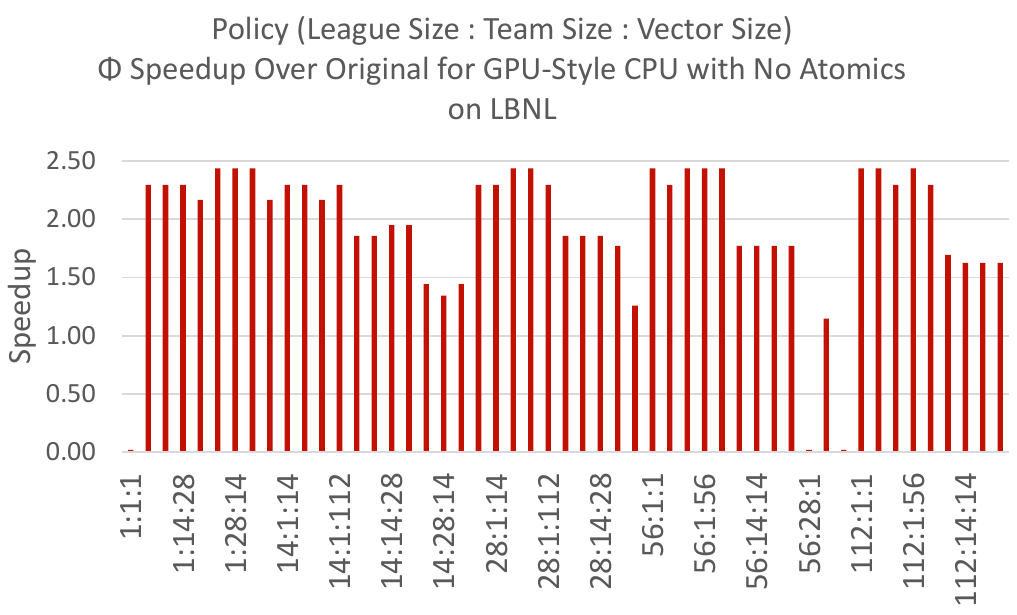}
\caption{Fine parallel policy parameter grid search for overall SparTen CP-APR MU and $\bm\Phi^{(n)}$  computation on CPU for {\it LBNL}, varying league size, team size, and vector size.}
\label{fig_policy_cpu_lbnl}
\end{figure}

\begin{figure}[!t]
\centering
\includegraphics[width=0.8\textwidth]{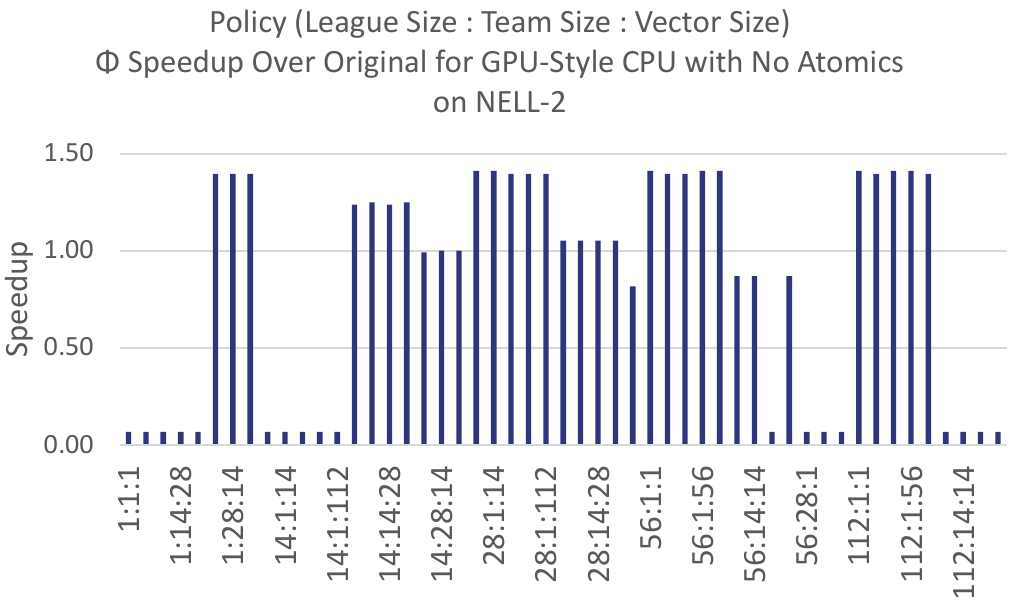}
\caption{Fine parallel policy parameter grid search for overall SparTen CP-APR MU and $\bm\Phi^{(n)}$ computation on CPU for {\it NELL-2}, varying league size, team size, and vector size.}
\label{fig_policy_cpu_nell_2}
\end{figure}

For {\it Chicago}, eliminating atomic operations allows the $\bm\Phi^{(n)}$ computation to achieve speedups ranging from $0.04\times$ to $1.97\times$, and full CP-APR MU to achieve speedups ranging from $0.0\times$ to $2.0\times$, compared to the SparTen CPU baseline.
For {\it LBNL}, $\bm\Phi^{(n)}$ speedups ranged from $0.0\times$ to $2.0\times$, and full CP-APR MU speedups ranged from $0.0\times$ to $2.4\times$ compared to the baseline.
Finally, for {\it NELL-2}, $\bm\Phi^{(n)}$ speedups ranged from $0.1\times$ to $1.7\times$, and full CP-APR MU speedups ranged from $0.1\times$ to $1.4\times$ compared to the baseline.
The results in Figures~\ref{fig_policy_cpu_chicago}--\ref{fig_policy_cpu_nell_2} show that a better Kokkos policy can further improve the performance of the GPU implementation running on CPU;
however, as discussed above, these policies must be chosen carefully to avoid significant degradation in performance, suggesting that having a good heuristic is important.

\clearpage

\subsection{Experiment 6: Kokkos Policy Study Across Modes of the Tensors}
We have so far considered the performance of the $\bm\Phi^{(n)}$ kernel for each iteration of the algorithm as a whole.
However, $\bm\Phi^{(n)}$ kernel computation is executed for every mode of the tensor within the iteration, and the performance of $\bm\Phi^{(n)}$ computation for each mode may be drastically different, given that the sparsity pattern (i.e., the data access pattern) typically changes across the modes.
In this experiment we examine the $\bm\Phi^{(n)}$ kernel performance across different modes.

\begin{figure}[!t]
\centering
\includegraphics[width=0.8\textwidth]{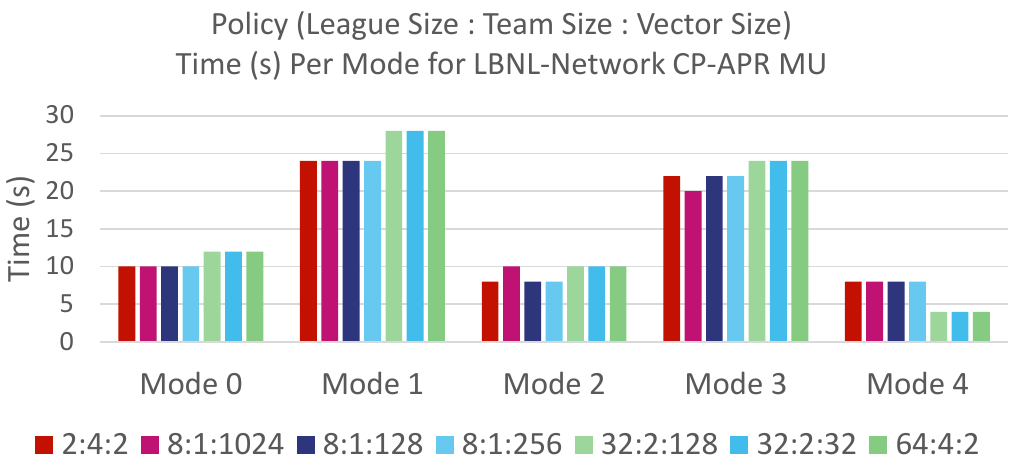}
\caption{%
Execution time for policy parameter grid search on SparTen CP-APR MU $\bm\Phi^{(n)}$ computation on CPU for {\it LBNL} for different modes and policy configurations.
}
\label{fig_by_mode_cpu_lbnl}
\end{figure}

\begin{figure}[!t]
\centering
\includegraphics[width=0.8\textwidth]{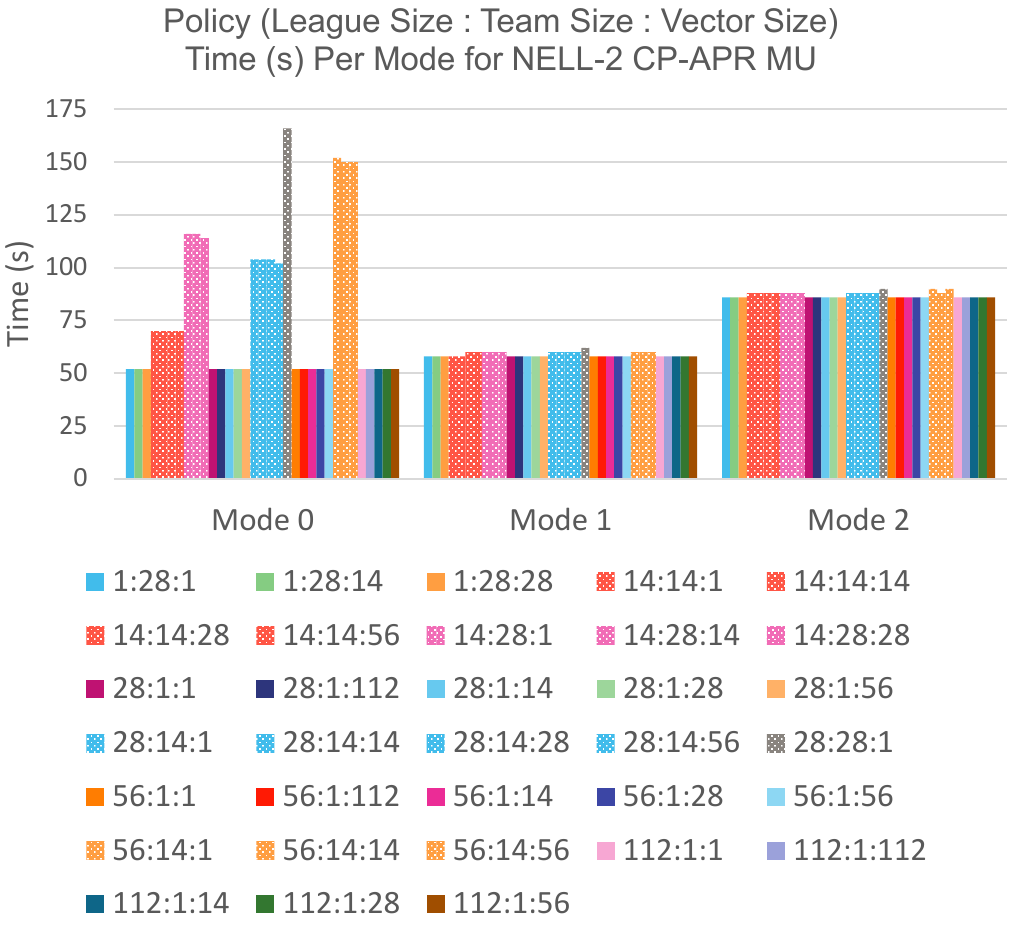}
\caption{By-mode timing results for policy parameter grid search on SparTen CP-APR MU $\bm\Phi^{(n)}$ computation on CPU for  {\it NELL-2}, varying league size, team size, and vector size.}
\label{fig_by_mode_cpu_nell_2}
\end{figure}

We evaluate the SparTen CPU implementation on two input tensors with a coarse-grained grid search over the Kokkos parallel policy. 
For {\it LBNL}, we vary league size from $2$ to $64$, team size from $1$ to $4$, and vector size from $2$ to $1024$.
For {\it NELL-2}, we vary league size from $1$ to $112$, team size from $1$ to $28$, and vector size from $1$ to $56$. 
Any invalid configurations (those violating the Kokkos requirement that $\textrm{team size} \times \textrm{vector size} \leq 1024$) are omitted.
Results are shown in Figures~\ref{fig_by_mode_cpu_lbnl} and~\ref{fig_by_mode_cpu_nell_2} for {\it LBNL} and {\it NELL-2}, respectively.
Performance for {\it LBNL} is relatively consistent across different modes, while {\it NELL-2} exhibits distinct anomalies for the first mode where performance suffers significantly for specific configurations.
This result further suggests that a good heuristic for selecting Kokkos policy is essential in maintaining performance portability.

\clearpage

\subsection{Experiment 7: Tensor Operation Performance in the STREAM Benchmark}
\label{sec:stream}

In the final two studies, we take a top-down approach and decompose the $\bm\Phi^{(n)}$ computation to more fundamental operations and analyze their performance.
This allows us to take a more fine-grained approach to determining the fundamental operational bottlenecks in the CP-APR MU algorithm.
We describe these fundamental tensor operations in Table~\ref{tab:stream}, along with their operational intensity, $I$.
Note that these are operations are supported by the well-known \emph{STREAM} benchmark~\cite{stream1995}, which we extended with a Kokkos version.

\begin{table}[!htbp]
\centering
    \caption{Fundamental tensor operations}
        \begin{tabular}{l|l|c|c|c} 
         Name & Kernel & Bytes & Ops & $I$\\
         \hline
         \hline
            Copy    & A[i] = B[i]               & 16 & 0 & 0\\
            Scale   & A[i] = s * B[i]           & 16 & 1 & 0.0625\\
            Add     & A[i] = B[i] + C[i]        & 24 & 1 & 0.042\\
            Triad   & A[i] = B[i] + s * C[i]    & 24 & 2 & 0.083\\
         \hline
        \end{tabular}
\label{tab:stream}
\end{table}
\noindent


Implementing Kokkos parallel constructs within an existing code base is a straightforward process of refactoring only targeted code regions to utilize the parallel code execution and data management in the Kokkos programming model. 
We first identify parallel regions in the code, such as those within existing OpenMP {\tt \#pragma} statements, and replace them with Kokkos \texttt{parallel\_for} dispatch while incorporating the loop body into a C++ lambda expression. (Note that OpenMP 4.5+ supports offloading to GPU devices~\cite{openmp}, but we use Kokkos for performance portability due to its ability to efficiently handle data layout for both dense and sparse operations.) 
The next step is to refactor nested parallel regions and to store data in Kokkos abstractions called \emph{Views}, after which the code is completely portable to any Kokkos-supported hardware platform backend. 
Nested parallel regions map to SIMD instructions when compiling with Kokkos for CPU and to thread blocks for GPU targets. Note that while \emph{STREAM} kernels do not require nested parallel regions, we briefly investigated employing them in these kernels and found the performance to be lower than that of a single-level parallel region. Results using the single-level parallel region approach for \emph{STREAM} are what we present here.

\begin{figure}[!htb]
\centering
\includegraphics[width=0.725\textwidth]{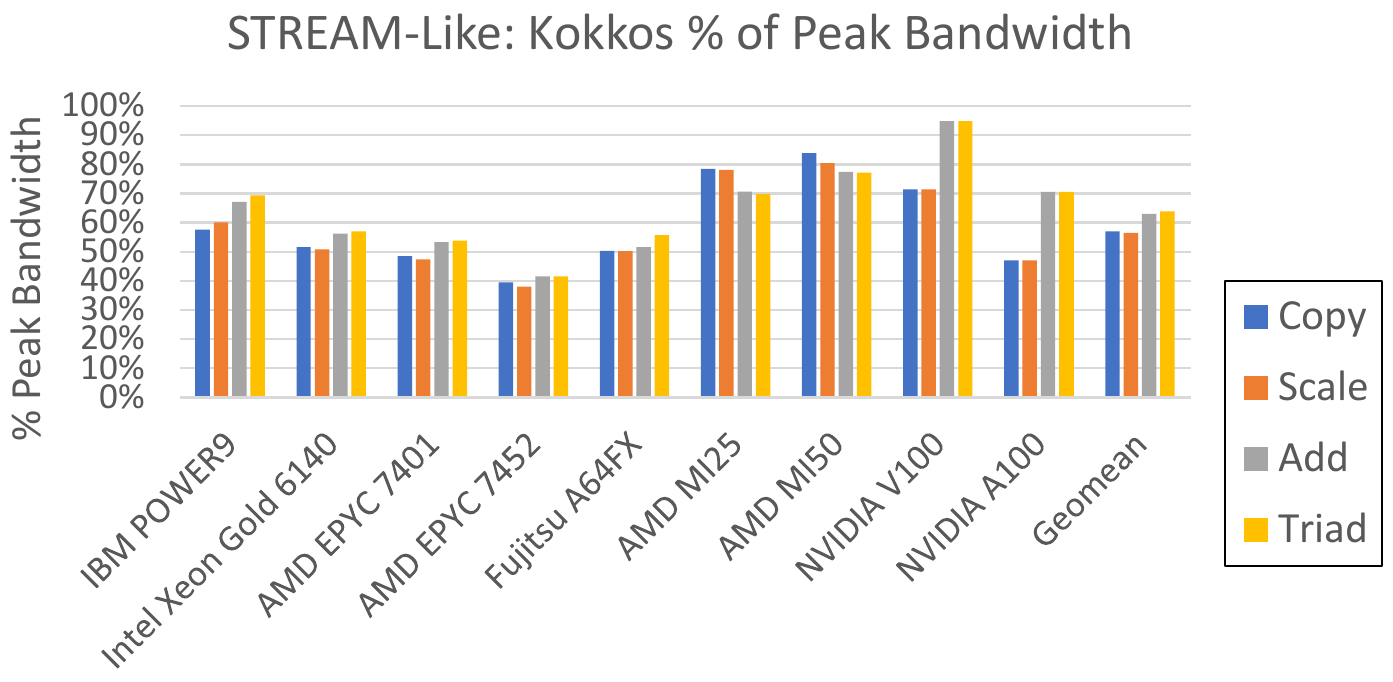}
\caption{Kokkos percentage of system peak obtained with the STREAM-like benchmark.}
\label{fig_stream_like_peak}
\end{figure}

\begin{figure}[!htb]
\centering
\includegraphics[width=0.725\textwidth]{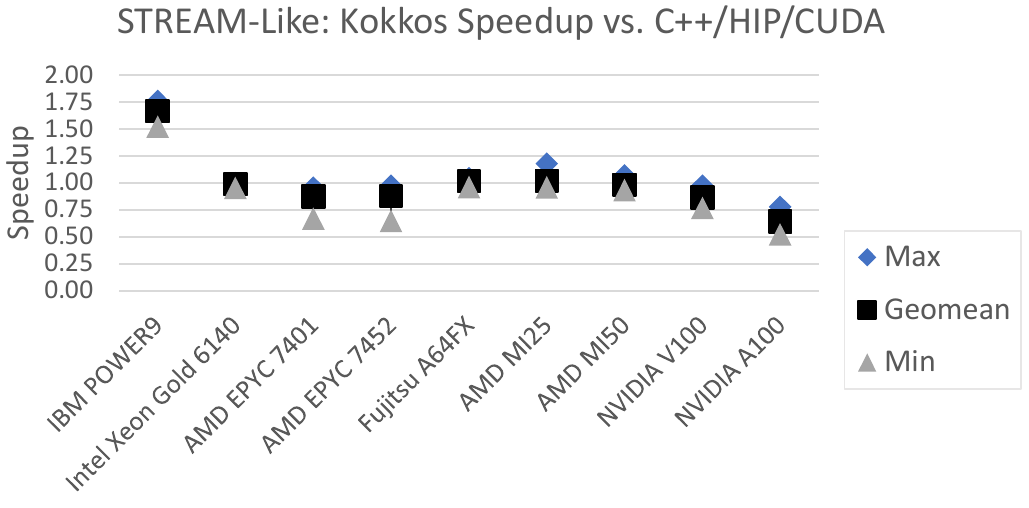}
\caption{Kokkos speedup over hand-tuned code for the STREAM-like benchmark.}
\label{fig_stream_like_speedup}
\end{figure}

We proceed by porting the simple \emph{STREAM} kernels to Kokkos, then measuring the bandwidth.
We evaluate our ported code on a wider set of test systems than used in the previous experiments presented above. Table~\ref{tab:newsystems} presents the full list of CPU and GPU systems we used in our experiments.
Figure~\ref{fig_stream_like_peak} shows the measured bandwidth as a percentage of peak theoretical system bandwidth.
We also compare the performance of our Kokkos-enhanced \emph{STREAM} implementation against the original \emph{STREAM} benchmark by taking the geometric mean speedup over the four operations on each system, as shown in Figure~\ref{fig_stream_like_speedup}.


From this experiment we can see that Kokkos achieves on average approximately $50\%$ of the system peak memory bandwidth over the test systems, and that this is largely on par with the original \emph{STREAM} implementation, with the notable exceptions of the IBM POWER9 CPU, where Kokkos achieves a geometric mean $1.66\times$ speedup, and the NVIDIA A100 GPU, where Kokkos exhibits a $0.64\times$ slowdown.
The original \emph{STREAM} benchmark uses OpenMP and runs on CPUs only. 
For GPU \emph{STREAM} results we used the GPU-STREAM benchmark which has HIP/AMD and CUDA/NVIDIA implementations.  The term ``STREAM-like'' in the figures is used per the original \emph{STREAM} author guidelines to distinguish our Kokkos implementation and GPU-STREAM from the original \emph{STREAM} benchmark proper.)


\begin{table}[!htbp]
\centering
\caption{Test systems for fundamental tensor operation evaluation}
        \begin{tabular}{c|c|c} 
            Type & Name & \# Cores \\ 
            \hline
            \hline
            CPU & IBM POWER9 & 20 \\ 
            CPU & Intel Xeon Gold 6140 & 18 $\times$ 2 (dual-socket)\\ 
            CPU & AMD EPYC 7401 & 24 $\times$ 2 (dual-socket)\\ 
            CPU & AMD EPYC 7452 & 32 $\times$ 2 (dual-socket)\\ 
            CPU & Fujitsu A64FX & 48\\ 
            \hline
            GPU & AMD Vega MI25 & 4096 (STREAM Processors) \\ 
            GPU & AMD Vega MI50 & 3840 (STREAM Processors) \\ 
            GPU & NVIDIA V100 & 5120 (CUDA Cores) \\
            GPU & NVIDIA A100 & 6912 (CUDA Cores) \\ 
            \hline
        \end{tabular}
        \label{tab:systems}
\label{tab:newsystems}
\end{table}

\subsection{Experiment 8: MTTKRP Operations in the PASTA Benchmark}

Another important operation in tensor decomposition is the matricized tensor times Khatri-Rao product (MTTKRP), which is the primary bottleneck for the CP-ALS tensor decomposition algorithm (see~\cite{KoBa09} for details and references therein). MTTKRP computation is characterized by the following operations:
\begin{align}
    T(:) & \leftarrow B(j,:) * C(k,:) & \textrm{element-wise product}\\
    T(:) & \leftarrow v * T(:) & \textrm{scale}\\
    A(i,:) & \leftarrow A(i,:) + T(:) & \textrm{element-wise add}\label{eq:accum}
\end{align}

We begin with the MTTKRP kernel from the Parallel Sparse Tensor Algorithm Benchmark Suite ({\it PASTA}) library~\cite{Li:2019:PASTA} and port it to Kokkos as described in Section~\ref{sec:stream}. We then compare system peak bandwidth using our Kokkos version {\it PASTA} MTTKRP reference version. We use the systems described in Table~\ref{tab:newsystems} and the following subsets of tensors from the FROSTT dataset: {\it Chicago}, {\it NELL-2}, {\it NIPS}, and {\it Uber}. Note that unlike the simpler \emph{STREAM} kernels, the MTTKRP algorithm lends itself to using Kokkos nested parallelism.

\begin{figure}[!hbp]
\centering
\includegraphics[width=0.8\textwidth]{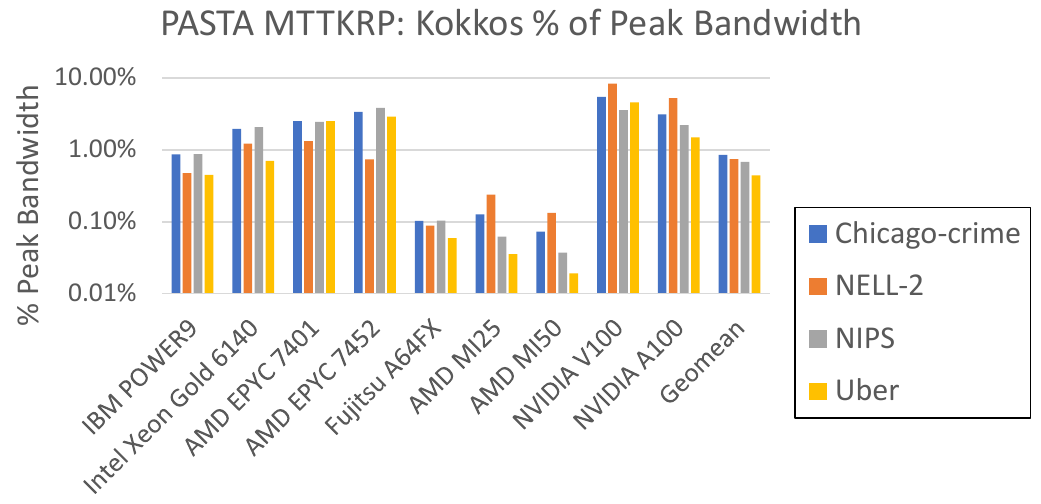}
\caption{Kokkos percentage of system peak obtained with the PASTA-like MTTKRP benchmark.}
\label{fig_pasta_like_peak}
\end{figure}

\begin{figure}[!hbp]
\centering
\includegraphics[width=0.8\textwidth]{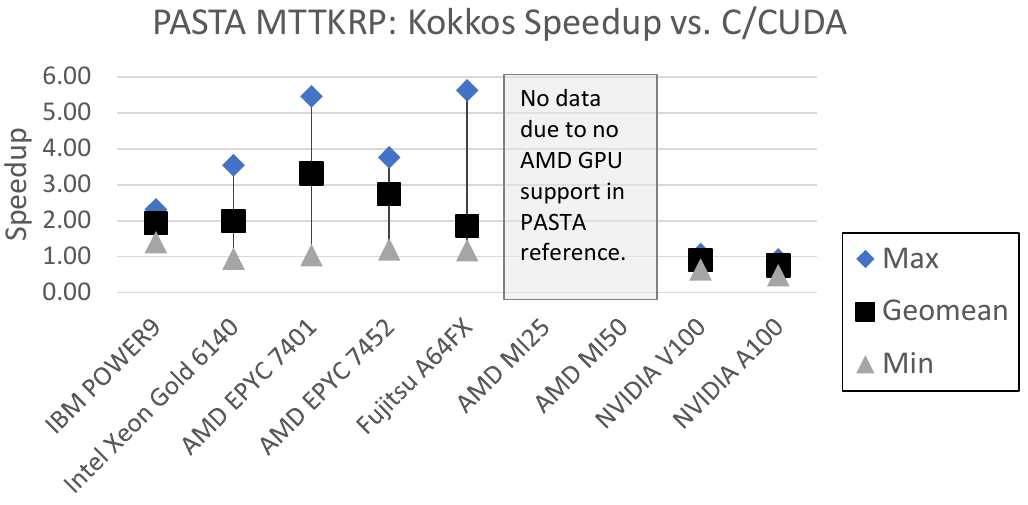}
\caption{Kokkos speedup over hand-tuned code for the PASTA-like MTTKRP benchmark.}
\label{fig_pasta_like_speedup}
\end{figure}

The results are shown in Figures~\ref{fig_pasta_like_peak} and~\ref{fig_pasta_like_speedup}. 
We can see from Figure~\ref{fig_pasta_like_speedup} that the Kokkos implementation achieves significant speedup over PASTA on most systems.
We also achieve a very low percentage of theoretical peak memory bandwidth, but this is likely due to the memory load/store bottleneck in the MTTKRP kernel~\cite{8425210} that makes the kernel latency-bound, as we perform on par or better than the state-of-the-art {\it PASTA} benchmark.
As in the \emph{STREAM} benchmark experiment, Kokkos exhibits a slowdown on the NVIDIA A100 GPU ($0.76\times$) when compared to the {\it PASTA} CUDA MTTKRP baseline. However, Kokkos achieves geometric mean speedups of $1.85\times$ to $3.32\times$ on the CPU systems; The large variance for the CPU systems is due mostly to the performance of {\it NELL-2}, which achieves the maximum speedup on Fujitsu A64FX ($5.63\times$ speedup, also the overall CPU maximum speedup) and achieves the maximum slowdown on Intel Xeon Gold 6140 ($0.94\times$ slowdown, also the overall CPU maximum slowdown). On the other CPU systems, {\it NELL-2} achieves the minimum speedup of the tensors tested and is notably above $1.0\times$ speedup on those systems. This result shows that Kokkos offers good performance portability on CPUs, and furthermore has an advantage for {\it NELL-2}-like data on Fujitsu~A64X-style processors.
Note that {\it PASTA} does not support AMD GPUs, so there are no speedup results for the AMD GPUs. Additionally, {\it PASTA} at time of writing supports only 3-way and 4-way tensors on GPU.

\clearpage

\section{Discussion and Future Work}
\label{sec:conc}

In this report, we present analysis of the CP-APR MU tensor decomposition algorithm using the Roofline Model and Pressure Point Analysis (PPA).
We used these analysis techniques to determine that 
\begin{itemize}
    \item $\bm\Phi^{(n)}$ computation is the most time consuming portion of the SparTen CP-APR MU implementation,
    \item $\bm\Phi^{(n)}$computation performance is limited by the memory bandwidth (via the Roofline Model),
    \item atomic operations are not a critical bottleneck and enable higher performance on GPUs due to their caching mechanism (via PPA), and
    \item higher data reuse in cache will provide non-trivial improvements in performance for the $\bm\Phi^{(n)}$ computation (via PPA).
\end{itemize}
Additionally, we conducted an extensive Kokkos policy evaluation to determine that further performance gains can be observed through manual kernel launch parameter tuning compared to the automatic, default Kokkos policy. Moreover, a poor choice of the Kokkos policy may lead to significant performance degradation.
This suggests that the development of a heuristic for determining the optimal Kokkos policy for a given computation could increase the performance portability of Kokkos.

Our top-down study on the fundamental tensor operations and kernels suggests that for both simple and complex operations and kernels, Kokkos achieves comparable performance to state-of-the-art benchmarks such as \emph{STREAM} and \emph{PASTA}.
Overall, our evaluation suggests that Kokkos demonstrates good performance portability for simple operations (e.g., \emph{STREAM}) and algorithms that consist of a sequence of simple operations (e.g., \emph{PASTA}), but requires architecture-specific tuning and scheduling for algorithms with more complex dependencies and data access patterns.
One obvious next step for Kokkos is to design a practical work-thread mapping and scheduling heuristic.
\addcontentsline{toc}{section}{References}
\bibliography{arxiv-refs}

\end{document}